\documentclass[twocolumn,prd,aps,superscriptaddress]{revtex4}

\usepackage{graphicx}% Include figure files
\usepackage{dcolumn}% Align table columns on decimal point
\usepackage{bm}% bold math

\begin{document}
%\topmargin = 0 true in

\title{Exotics and all that}

\author{T. Barnes}
\affiliation{Physics Division, Oak Ridge National Laboratory, Oak Ridge,
TN 37831}
\affiliation{Department of Physics and Astronomy, 
University of Tennessee, Knoxville, TN
37996}

\begin{abstract}
\vskip .3 true cm
This invited contribution summarizes some of the more important
aspects of exotics. We
review theoretical expectations for exotic and nonexotic
hybrid mesons, and briefly
discuss
the leading experimental candidate for an exotic,
the $\pi_1(1600)$.
%
%\PACS{
%      {12.39}{quark models}   \and
%      {12.40.Y}{hadron mass models}  \and
%      {13.25}{hadronic decays of mesons}
%     } % end of PACS codes
%} %end of abstract
\end{abstract}

\maketitle

\section{Introduction}

\label{intro}

An ``exotic meson'' is a meson resonance with
J$^{PC}$ or flavor quantum numbers that are forbidden
to
$|q\bar q\rangle $.

The experimental candidates of current interest are
``spin-parity exotics'', which have J$^{PC}$ forbidden to
$q\bar q$ mesons. In principle one might also find flavor exotics
in a multiquark sector, for
example in I=2, but no
widely accepted experimental candidates are known at present~\cite{GN}.

Every physical meson is actually
a linear superposition of all allowed
basis states, spanning (unless strictly forbidden)
$|q\bar q  \rangle$, $|q\bar q g  \rangle$,
$|q^2\bar q^2\rangle$,
$|gg  \rangle$ and so forth,
with amplitudes that
are determined by QCD interactions.
Our working
classification of resonances as ``quarkonia'', ``hybrids'', ``glueballs''
and so forth is a convenience that
implicitly assumes that one type of basis state dominates the state
expansion of each resonance. Of course this may not be the case in general,
and the amount of ``configuration mixing'' is an important and little-studied
topic in hadron physics \cite{Bur02}.
Exotics are the special cases
in which the $|q\bar q\rangle$ component {\it must} be zero,
due to the quantum numbers of the resonance.

\section{Hybrids}
\label{sec:1}
A hybrid meson is a resonance whose
dominant valence component is $|\, q \bar q + {\hbox {\it excited glue}}
\rangle$.
At present this is a somewhat imprecise
and model-dependent definition, as there are
several dissimilar models of
the nature of gluonic excitations in the low energy,
nonperturbative regime.
The best known of these descriptions of excited glue are the
bag model,
constituent-gluon models,
and the flux-tube model.
Fortunately these very different intuitive pictures of excited glue
lead to rather similar predictions for the masses and quantum numbers
of low-lying hybrids.

One general conclusion is that unlike $q\bar q$,
all J$^{PC}$ quantum numbers are spanned
by hybrids. This can be seen either by considering
explicit models or through
an enumeration of all
$ (\bar \psi\Gamma \lambda^a/2 \psi ) \otimes F^a$ interpolating
fields. The list of allowed hybrid
quantum numbers thus includes J$^{PC}$ combinations forbidden to
$q\bar q$ states, which are called ``spin-parity exotics";
\begin{displaymath}
{\rm J}^{PC}\bigg|_{exotic} =
0^{--}, 
0^{+-}, 
1^{-+}, 
2^{+-}, 
3^{-+}, \dots \ .
\end{displaymath}
These quantum numbers are extremely attractive experimentally,
since such resonances cannot be ordinary
$q\bar q$ quark model states.

\subsection{Models of hybrids}
\label{sec:2}

\subsubsection{Introduction}
\label{sec:3}

Much of the work on
hybrids has made use of specific models of
``excited glue'', especially the bag model,
constituent-gluon, and flux-tube model.
J$^{PC}$-exotic hybrids
may also be studied without specializing to a model through
QCD sum rules and LGT.
We will summarize some of the
predictions of these models and techniques, specifically
the mass spectrum,
quantum numbers and decay properties.

\subsubsection{Bag model}
\label{sec:4}

The bag model assumes a spherical hadron, with quarks and gluons
populating cavity modes that are confined by boundary conditions
on the colored quark (Dirac) and gluon (Maxwell) fields.
The ``zeroth-order'' bag model basis states are
color-singlet products of quarks, antiquarks and
gluons occupying cavity modes,
\begin{displaymath}
|q\bar q\rangle \ , \hskip .3cm 
|q\bar qg \rangle \ , \hskip .3cm
|gg \rangle \ , \hskip .3cm
|q^2\bar q^2 \rangle \ , \dots \ .
\end{displaymath}
The quark-gluon and gluon self-interactions of QCD
mix these basis states,
so the physical levels are linear superpositions.

Combining the lowest-lying $q, \bar q$ and (J$^P = 1^+$) $g$ modes, one
finds the lowest bag-model hybrid multiplet
\begin{displaymath}
{\rm J}^{PC}\big|_{\rm bag\ model}   =
( 0^-, 
1^- 
) \otimes 1^+ = 
1^{--}; 
0^{-+}, 
1^{-+}, 
2^{-+}\ .
\end{displaymath}
The $1^{-+}$ combination is a ${\rm J}^{PC}$-exotic. Without
the incorporation of pQCD quark-gluon and gluon-gluon interactions
these levels would be degenerate.
Detailed calculations of configuration
mixing between these quark+gluon basis states by pQCD
interactions finds
the level ordering $0^{-+} < 1^{-+} < 1^{--} < 2^{-+}$,
with a total multiplet splitting of {\it ca.} 500~MeV, and an
exotic mass of $M(1^{-+})\approx~1.5$ GeV \cite{BagH}.
As each of these J$^{PC}$ levels
is a flavor nonet in the $u,d,s$ system,
the bag model predicts many exotic and nonexotic
hybrids
at relatively low masses
that might be experimentally accessible.

\subsubsection{Flux-tube model}
\label{sec:5}

In LGT simulations a roughly cylindrical
region of modified
glue fields can be observed between widely separated static color sources
\cite{Bal01}.
This ``flux tube'' is the origin of the confining
linear potential between
$q$ and $\bar q$
in a color-singlet $q\bar q$ meson.
The flux-tube model \cite{ft_model} is an approximate description
of this state of glue, which is treated as a string of point masses,
``beads'', connected
by a linear potential.
This system supports locally transverse excitations that are treated
quantum mechanically by solving the bead Schr\"odinger equation.
The orbital angular momentum carried by this model flux tube
is combined with the $q\bar q$ spin and
orbital angular momentum to form
states of definite overall J$^{PC}$.
The lowest flux-tube hybrid multiplet spans
8 J$^{PC}$ levels,
with
\begin{displaymath}
{\rm J}^{PC}\bigg|_{flux-tube\ hybrids}  =
1^{\pm\pm};
0^{\pm\mp}, 
1^{\pm\mp}, 
2^{\pm\mp} 
\ .
\end{displaymath}
The first 2
levels have $S_{q\bar q}=0$ and the
remainder have $S_{q\bar q}=1$. Note that
this is a doubling of the bag model states listed earlier,
with the second set having
the opposite $(C,P)$.
These states are all degenerate in the simplest version of the
flux-tube model.

This model typically finds rather higher hybrid masses than
the bag model.
Isgur, Kokoski and Paton \cite{IKP} used small-oscillation
and adiabatic approximations
and found the lightest hybrid multiplet at
1.9(1)~GeV; a subsequent Hamiltonian Monte Carlo study
\cite{BCS}
found a very similar mass of 1.8-1.9~GeV
for this multiplet.

Since each of these 8 J$^{PC}$ levels spans a flavor nonet,
the flux-tube model predicts a very rich spectrum, with 72
meson resonances
expected
in the vicinity of 1.9~GeV,
in addition to the conventional $q\bar q$ quark-model states!

\subsubsection{LGT and QCD Sum Rules}
\label{sec:6}

LGT and QCD sum rules estimate masses
by evaluating the correlation functions
$\langle 0|{\cal O}(\vec x,\tau) {\cal O}^\dagger(0,0)|0 \rangle $, where
$ {\cal O}^\dagger$ is an operator that
excites the state of interest
from the vacuum. This approach uses the fact that
these correlation functions summed over $\vec x$
at large $\tau$
approach an exponential in the mass
of the lightest state excited by the operator
${\cal O}^\dagger$.

Both of these methods have systematic errors.
QCD
sum rules relate these correlation functions
to pQCD contributions and
VEVs of other operators that are inferred from
experiment, and different choices for these VEVs
and uncertainties
in higher-mass contributions (and algebra errors)
have led to a wide
scatter of results in the literature.
QCD sum rule
mass estimates for the light $1^{-+}$ exotic
range from $\approx 1$~GeV to
2.1~GeV, with the higher masses preferred by the more recent
references \cite{QCDsrexotic}.
A few other exotics have been studied using
QCD sum rules; the $0^{--}$ for example
has been found to have
a rather
high mass of {\it ca.} 3 GeV.

LGT results for $0^{++}$ glueballs and
$1^{-+}$ exotic hybrids were recently reviewed by
McNeile \cite{McN02a}, and a detailed review of the approach
has been published by Bali \cite{Bal01}.
Most LGT studies to date have used the ``quenched approximation",
which neglects
the effects of coupling to decay channels. Unfortunately these
effects may include important mass shifts.
Exotic hybrid masses have been studied by several groups, recently
including  the MILC
collaboration \cite{MILC}
(light
$1^{-+}$ and $0^{+-}$ exotics),
UKQCD
\cite{UKQCD}
($0^{+-}$, $1^{-+}$ and $2^{+-}$;
these are the three exotics
predicted to be lightest, and degenerate, in the
zeroth-order flux-tube model), and
Luo and Mei \cite{Luo02}
(light and $c\bar c$ $1^{-+}$).

Recent LGT
results are approximately consistent with
the flux-tube model; signals in all
three low-lying flux-tube exotic
channels are observed, with the mass of the $1^{-+}$
(the best determined)
being about 2.0~GeV. The $0^{+-}$ and $2^{+-}$ may lie somewhat higher,
but this is unclear with present statistics.

The application of
LGT
to nonrelativistic heavy quark systems has been of much
recent interest. Considerably
reduced statistical errors follow from the use of an
``NRQCD" action derived from a heavy quark expansion.
This approach has been applied
to  $1^{-+}$ heavy-quark exotic hybrids;
the $1^{-+}$ $b\bar b$ hybrid is found to lie near
$11.0$ GeV, and the
$1^{-+}$ charmonium hybrid
is predicted to lie just below $4.4$ GeV. (See Ref.\cite{Luo02} for
a summary.)
These LGT results strongly motivate a high-statistics
scan of $R$ near these masses, since models of hybrids anticipate that the
multiplet containing the
$1^{-+}$ will also possess a $1^{--}$ hybrid nearby in mass.

\subsection{Hybrid baryons}
\label{sec:7}

One may also form hybrid baryon from $qqq$ and excited glue.
Bag model calculations
\cite{Hbar}
predict a lowest multiplet of
$u,d$ hybrid baryons with
(J$^P$, flavor)
$(1/2^+ {\rm N})^2$,  $(3/2^+ {\rm N})^2$,
$(5/2^+ {\rm N})$,
$(1/2^+ \Delta)$,
$(3/2^+ \Delta)$.
Calculations of configuration mixing through
quark-gluon and gluon-gluon interactions
predict a rather large
overall multiplet splitting of {\it ca.} 500~MeV. The resulting
lowest-lying hybrid baryon is found to be a
$(1/2^+ {\rm N})$ level, with a mass near 1.5~GeV.
Recent flux-tube model calculations of hybrid baryons \cite{CapP}
find a rather similar spectrum of low-lying states,
starting with degenerate
$(1/2^+ {\rm N})^2$ and  $(3/2^+ {\rm N})^2$ states
at 1870(100)~MeV, followed by
$(1/2^+ \Delta)$,
$(3/2^+ \Delta)$ and
$(5/2^+ \Delta)$.
Unfortunately there are no baryon J$^P$-exotics, so searches for these
levels must establish an overpopulation of experimental baryons
relative to the theoretical $qqq$ quark model spectrum.

\section{Hybrid decays}
\label{sec:8}

There is general theoretical
agreement that hybrid resonances exist, and that
the lightest $u,d$ hybrid meson multiplet includes a $1^{-+}$
resonance with a mass in the $1.5$-$2$ GeV region.
Theoretical predictions of the strong decay widths of these states
are of great interest, since
many otherwise
experimentally
attractive
decay channels may have weak couplings to hybrids, or may
couple dominantly to hybrids that are
so broad as to
be difficult to identify, a problem familiar from the $f_0$ sector.

Several strong decay models been applied to hybrids.
The best known is the flux-tube decay model, which was
applied to exotic hybrids by Isgur, Kokoski and Paton \cite{IKP} and
subsequently to nonexotic hybrids by Close and
Page \cite{CP}.
This model assumes that decays take place by
$^3$P$_0$
$q\bar q$ pair production
along the length of the flux tube. For
the {\it unexcited} flux tubes of conventional mesons the predictions are
quite similar to the rather successful $^3$P$_0$ model; for hybrids
this decay model leads to predictions of very characteristic strong
decay amplitudes.

In the flux-tube decay model the
orbital angular momentum of the hybrid's excited flux tube
gives the $q\bar q$ source
produced in the decay a phase dependence around the
axis of the original $q\bar q$,
and the hadronic final states produced most strongly
are those which have similar
angular dependence.
As a result, many of the well-studied simple final states
such as $\pi\pi$, $\rho\pi$ and so forth are predicted
to be produced quite weakly in hybrid decays,
due to poor spatial overlap with this
$e^{i\phi}$-dependent $q\bar q$ source. The favored modes are those that
have a large L$_z=1$ axial projection, such as an S+P meson pair. This is the
origin of the flux-tube S+P
selection rule, which in the I=1  $1^{-+}$ case favors
the unusual modes $\pi f_1$ and $\pi b_1$ over
$\eta\pi$, $\eta'\pi$ and $\rho\pi$, despite their more limited
phase space. Caution is appropriate here, since recent
studies of the decay modes of orbitally-excited
quarkonia in the $^3$P$_0$ model also find
a preference for S+P modes in many cases \cite{3P0}.

Hybrid strong decays have also been studied using QCD
sum rules \cite{QCDsrexotic}, a vector flux-tube model
\cite{PSS}, and constituent-gluon models \cite{cgdecay}.
There is agreement (with some variation between models)
that in most cases
S+P modes dominate hybrid strong decays.

Due to the difficulty of treating strong decays on the lattice
there have been few studies of this very important subject.
One recent, very interesting result is a LGT study of closed-flavor
strong decays of heavy-quark hybrids, by McNeile, Michael and
Pennanen \cite{McN02b}. This work finds that hybrid strong
decays of the
type $1^{-+}\to \chi S$, where $\chi$ is a P-wave $Q\bar Q$ and
$S$ is a light $q\bar q$ scalar, are much larger than expected;
partial widths in the 10s of MeV appear likely. This is excellent
news for experimental searches, since transitions such as
$H_c \to \chi_c S \to \gamma J/\psi (\pi \pi)_S$,
$J/\psi \to \ell^+ \ell^-$ allow efficient background
rejection.
Closed-flavor strong cascades had been suggested
previously as a method for searching for heavy hybrids
in $e^+e^-$ annihilation
(see for example Ref.\cite{TCF95}), but it was
thought that the transition rates would be much smaller.
If the new LGT results are correct, this approach now
appears very attractive.

\section{An experimental exotic meson: $\pi_1(1600)$}
\label{sec:9}

At present there are just two experimental candidates
for exotic mesons,
the $\pi_1(1400)$ and the $\pi_1(1600)$. In view of the limited space
available here, I
will only discuss the well established
$\pi_1(1600)$. The long and complicated history of the $\pi_1(1400)$
is summarized elsewhere \cite{Bar00}.

Evidence for the I=1, J$^{PC}=1^{-+}$ $\pi_1(1600)$ has been
reported in  three channels,
$b_1 \pi$ (VES \cite{VES_1600}),
$\eta'\pi$ (VES \cite{VES_1600} and E852 at BNL \cite{E852_etappi}) and
$\rho\pi$ (VES \cite{VES_1600} and E852 \cite{E852_rhopi}).
Clear resonant phase motion is seen relative to
the
well-known $q\bar q$ states
$a_2(1320)$ and $\pi_2(1670)$ in the
$\eta'\pi$ and $\rho\pi$ channels respectively.
The mass and width reported by VES and E852 are consistent,

\begin{equation}
M_{\pi_1} =
\cases{
{\rm 1.61(2)\ GeV}  
&{VES, all modes}
\cr
1.597\pm 0.010 {+0.045\atop -0.010}\ {\rm GeV}
&{E852, $\eta'\pi$}
\cr
1.593\pm 0.008 {+0.029\atop -0.047}\ {\rm GeV}
&{E852, $\rho\pi$},
}
\end{equation}

\begin{equation}
\Gamma_{\pi_1} =
\cases{
{\rm 0.29(3)\ GeV} 
&{VES, all modes}
\cr
0.340 \pm 0.040  \pm 0.050\  {\rm GeV} 
&{E852, $\eta'\pi$}
\cr
0.168 \pm 0.020  {+0.150\atop -0.012}\  {\rm GeV} 
&{E852, $\rho\pi$}.
}
\end{equation}
The $\pi_1(1600)$ signal is especially clear in $\eta'\pi$,
in part because $q\bar q$ states such as the
$a_2(1320)$ have small branching fractions to this channel. (See Fig.2
of Ref.\cite{E852_etappi}.)

The relative  $\pi_1(1600)$ branching fractions
reported by VES
for the final states
$b_1 \pi$,
$\eta'\pi$ and
$\rho\pi$
are

\begin{equation}
\Gamma(\pi_1(1600)\to f ) 
=
\cases{
\equiv 1 
&{$b_1\pi $}
\cr
1.0\pm 0.3
&{$\eta'\pi$}
\cr
1.6\pm 0.4
&{$\rho\pi$} \ .
}
\end{equation}

Although the $\pi_1(1600)$ is a well-established exotic resonance,
there are problems with identifying it with a hybrid.
One difficulty is the
$\approx 300$-$400$ MeV difference between flux-tube and LGT estimates
of $M\approx 1.9$-$2.0$ GeV and the $\pi_1(1600)$ mass. This
discrepancy might of course
be a result of the quenched approximation used in LGT.

A second problem with teh hybrid assignment
is that the
reported relative
branching fractions
are inconsistent with the predictions of the flux-tube model
that S+P modes should be dominant.
For this state the flux-tube model
predicts that
$b_1 \pi$ should be dominant,
with $\rho\pi$
weak and $\eta\pi$ and $\eta'\pi$ very small
\cite{IKP,CP}. Some $\rho\pi$ coupling
is expected in the flux-tube model due to different $\rho$ and $\pi$
spatial wavefunctions \cite{CP},
but this is expected to be a much smaller effect
in the $\eta\pi$ and $\eta'\pi$ modes.
Indeed, there is a generalized G-parity
argument
that these S+S partial widths would be zero except for
differences in the final spatial wavefunctions \cite{Page}.
Either these three modes are not all due to a
hybrid, or our
understanding of hybrid decays is inaccurate.

Future experimental studies of
the $\pi_1(1600)$ in {\it all} its
allowed strong decay modes
will be especially interesting as
tests of theoretical models of exotic
meson decays.
This
state is of special relevance for
the GlueX photoproduction facility
planned at Jefferson Laboratory \cite{AP,GlueX},
since the $\pi_1(1600)$ and
other resonances with significant
$\rho \pi$ couplings should be produced
copiously in
one-pion-exchange
photoproduction processes.
\section{Acknowledgments}
\label{sec:10}

This work was supported in part by
the United States Department
of Energy under contract DE-AC05-96OR22464 managed by UT Battelle at
Oak Ridge National Laboratory, and by the University of Tennessee.


\newpage

\begin{thebibliography}{99}

\bibitem{GN}
For a recent review of light meson spectroscopy
see
S.Godfrey and J.Napolitano,
Rev. Mod. Phys. 71, 1411 (1999), hep-ph/9811410.

\bibitem{Bur02}
T. Burch, K. Orginos and D. Toussaint,
Nucl. Phys. Proc. Suppl. 106, 382 (2002),
hep-lat/0110001.

\bibitem{BagH}
T.Barnes, PhD thesis, Caltech (1977);
Nucl. Phys. B158, 171 (1979);
T.Barnes and F.E.Close, Phys. Lett. 116B, 365 (1982);
M.Chanowitz and S.Sharpe, Nucl. Phys. B222, 211 (1983),
{\it err.} B228, 588 (1983);
T.Barnes, F.E.Close and F.deViron, Nucl. Phys. B224, 241 (1983).

\bibitem{Bal01}
G.S.Bali, K.Schilling and C.Schlichter,
Phys. Rev. D 51, 5165 (1995),
hep-lat/9409005. For a review of LGT applications to hadron
spectroscopy see
G.S.Bali,
Phys. Rept. 343, 1 (2001),
hep-ph/0001312.

\bibitem{ft_model}
N.Isgur and J.Paton, Phys. Rev. D31, 2910 (1985).

\bibitem{IKP}
N.Isgur, R.Kokoski and J.Paton, Phys. Rev. Lett. 54, 869 (1985).

\bibitem{BCS}
T.Barnes, F.E.Close and E.S.Swanson, Phys. Rev. D52, 5242 (1995).

\bibitem{QCDsrexotic}
I.I.Balitsky, D.Dyakanov and A.V.Yung,
Phys. Lett. B112, 71 (1982) quoted a
$1^{-+}$ mass of $\approx 1$~GeV. A subsequent paper,
Z. Phys. C33, 265 (1986), estimated a rather higher mass.
Other QCD sum rule results for the mass
of the $1^{-+}$ exotic are
J. Govaerts, F.deViron, D.Gusbin and J.Weyers,
Nucl. Phys. B248, 1 (1984), 1.3~GeV;
J.I Latorre, S.Narison, P.Pascual and R.Tarrach,
Phys. Lett. B147, 169 (1984),
1.7(1) GeV;
J.I.Latorre, P.Pascual and S.Narison,
Z. Phys. C34, 347 (1987),
2.1~GeV.
Most recently,
K.Chetyrkin and S.Narison,
Phys. Lett. B485, 145 (2000),
hep-ph/0003151v2 quote
$\approx 1.6$-$1.7$ GeV, with the radial hybrid only about
0.2 GeV higher in mass.
Chetyrkin and Narison
also consider
decay couplings; they find
that the $\pi_1(1600-1700)$ exotic
has
$\Gamma_{\pi\rho} \approx 300$~MeV,
but for
$\Gamma_{\pi\eta'}$ only $ \approx 3$~MeV.
This is inconsistent with the reported decays of both
experimental $\pi_1$ exotic candidates.

\bibitem{McN02a}
C.McNeile, hep-lat/0207001.

\bibitem{MILC}
C.Bernard {\it et al.} (MILC Collaboration),
Phys. Rev. D56, 7039 (1997).

\bibitem{UKQCD}
P.Lacock {\it et al.} (UKQCD Collaboration),
Phys. Lett. B401, 308 (1997).

\bibitem{Luo02}
X.-Q. Luo and Z.-H. Mei,
hep-lat/0209049.

\bibitem{Hbar}
T.Barnes and F.E.Close,
Phys. Lett. 123B, 89 (1983); Phys. Lett. 128B, 277 (1983);
C.E.Carlson, T.H.Hansson and C.Peterson,
Phys. Rev. D27, 1556 (1983);
E.Golowich, E.Haqq and G.Karl,
Phys. Rev. D28, 160 (1983); {\it err.} D33, 859 (1986).

\bibitem{CapP}
S.Capstick and P.R.Page,
Phys. Rev. D60, 111501 (1999), nucl-th/9904041.

\bibitem{CP}
F.E.Close and P.R.Page,
Nucl. Phys. B443, 233 (1995).

\bibitem{3P0}
For $(u,d)$ meson decays see
T.Barnes, F.E.Close, P.R.Page and E.S.Swanson,
Phys. Rev. D55, 4157 (1996),
hep-ph/9609339.
For decays of strange mesons see
T.Barnes, N.Black and P.R.Page,
nucl-th/0208072.

\bibitem{PSS}
P.R.Page, E.S.Swanson and A.P.Szczepaniak,
Phys. Rev. D59, 034016 (1999).

\bibitem{cgdecay}
M.Tanimoto, Phys. Lett. B116, 198 (1982);
A.LeYaouanc {\it et al.},
Z. Phys. C28, 309 (1985);
F.Iddir {\it et al.},
Phys. Lett. B207, 325 (1988);
B433, 125 (1998), hep-ph/9803470;
Yu.S.Kalashnikova,
Z. Phys. C62, 323 (1994).

\bibitem{McN02b}
C.McNeile,
Phys. Rev. D65, 094505 (2002),
hep-lat/0201006.

\bibitem{TCF95} T.Barnes,
{\it Charmonium Physics at a Tau-Charm Factory},
hep-ph/9308368.

\bibitem{Bar00}
T.Barnes,
Acta Phys. Polon. B31, 2545 (2000),
hep-ph/0007296.

\bibitem{VES_1600}
V.Dorofeev (VES Collaboration),
in Proceedings of WHS99, hep-ex/9905002.

\bibitem{E852_etappi}
E.I.Ivanov {\it et al.}
(E852 Collaboration),
Phys. Rev. Lett. 86, 3977 (2001).

\bibitem{E852_rhopi}
G.S.Adams {\it et al.}
(E852 Collaboration),
Phys. Rev. Lett. 81, 5760 (1998);
S.U.Chung {\it et al.},
Phys. Rev. D65, 072001 (2002).

\bibitem{Page}
P.R.Page, Phys. Lett. B402, 103 (1997).

\bibitem{AP}
A.Afanasev and P.R.Page,
Phys. Rev. D57, 6771 (1998).

\bibitem{GlueX}
A. R. Dzierba,
``QCD Confinement and the Hall D Project at Jefferson Lab",
in Proc. of {\it e+ e- Physics at Intermediate Energies},
(SLAC, Stanford, California, 30 Apr - 2 May 2001),
hep-ex/0106010.

\end{thebibliography}
\end{document}